\documentclass[12pt]{iopart}
\usepackage[top=3cm,bottom=3cm,left=3cm,right=3cm]{geometry}
\usepackage{lmodern}
\usepackage{microtype}
\usepackage{iopams}
\usepackage{fixmath}
\usepackage{braket}
\usepackage{color}
\usepackage{graphicx}
\usepackage{hyperref}
\usepackage[all]{hypcap}
\usepackage{footmisc}

\newcommand{\hc}{\ensuremath{\mathrm{h.c.}}}
\renewcommand{\vec}[1]{\mathbf{#1}}

\begin{document}
\title[Variational Monte Carlo phase diagram of the bilayer Hubbard model]
{Phase diagram of the square lattice bilayer Hubbard model: a variational Monte Carlo study}

\author{Robert R\"uger~\footnote{Present address: Scientific Computing \& Modelling NV, De Boelelaan~1083, 
1081~HV Amsterdam, The Netherlands.}, 
Luca F. Tocchio~\footnote{Present address: SISSA, via Bonomea~265, 34136~Trieste, Italy.}, 
Roser Valent\'\i{} and Claudius Gros}
\address{Institut f\"ur Theoretische Physik, Goethe-Universit\"at Frankfurt, Max-von-Laue-Stra{\ss}e~1, 
60438~Frankfurt am Main, Germany}
\ead{rueger@itp.uni-frankfurt.de}

\begin{abstract}
We investigate the phase diagram of the square lattice bilayer Hubbard model at half filling
with the variational Monte Carlo method for both the magnetic and the paramagnetic case as a function
of the interlayer hopping $t_\perp$ and on-site Coulomb repulsion $U$. With this study we resolve some discrepancies 
in previous calculations based on the dynamical mean field theory, and we are able to determine the nature of 
the phase transitions between metal, Mott insulator and band insulator. In the magnetic case we find only two 
phases: An antiferromagnetic Mott insulator at small $t_\perp$ for any value of $U$ and a band insulator 
at large $t_\perp$. 
At large $U$ values we approach the Heisenberg limit. The paramagnetic phase diagram shows at small $t_\perp$ 
 a metal to Mott insulator transition at moderate $U$ values and a Mott to band insulator transition 
at larger $U$ values.
We also observe a re-entrant Mott insulator to metal transition and metal to band insulator transition
for increasing $t_\perp$ in the range of $5.5t < U < 7.5t$. Finally, we discuss the obtained phase diagrams in 
relation to previous studies based on different many-body approaches.
\end{abstract}



\section{Introduction}

Understanding the origin of  transitions from a metal to a Mott or
a band insulator in correlated systems has been a topic of intensive debate
 in the past few years. Various generalizations of the Hubbard model
have been investigated for this purpose,
like the extended Hubbard model~\cite{PhysRevLett.99.216403}, the ionic Hubbard model in one and two
dimensions~\cite{PhysRevLett.92.246405, PhysRevLett.97.046403, PhysRevLett.98.016402,
PhysRevB.76.085112, 1367-2630-12-9-093021}, the two-band Hubbard model~\cite{Sentef2009} 
and the bilayer Hubbard
model~\cite{PhysRevB.73.245118, PhysRevB.75.193103, PhysRevB.77.144527, refId0, BilayerSpectra, PhysRevB.87.125141, 
PhysRevB.88.235115, PhysRevB.89.035139}. The latter has been specially important in relation
to the bilayer high-$T_c$ cuprates~\cite{Damascelli2003,Fournier2010}. Previous investigations of the above models
have been carried out primarily by employing dynamical
mean-field theory~(DMFT)~\cite{RevModPhys.68.13} and its
cluster extensions~\cite{DCA1, DCA2, DCA3, DCA4, CDMFT}.  While DMFT already captures a significant amount of
key properties in correlated systems, it is extremely important to analyse  these models with
completely unrelated many-body methods in order to get a deeper understanding of the underlying
physics.  In this work  we investigate  the phase diagram of the
square lattice bilayer Hubbard model at half filling with the variational Monte Carlo~(VMC) method.
 With this study 
(i) we are able to resolve some discrepancies between previous DMFT and cluster DMFT studies and (ii)
we find new aspects of  the Mott to band transition not captured in previous studies.

\begin{figure}
\centering
\raisebox{0.05\columnwidth}{
\includegraphics[height=0.25\columnwidth]{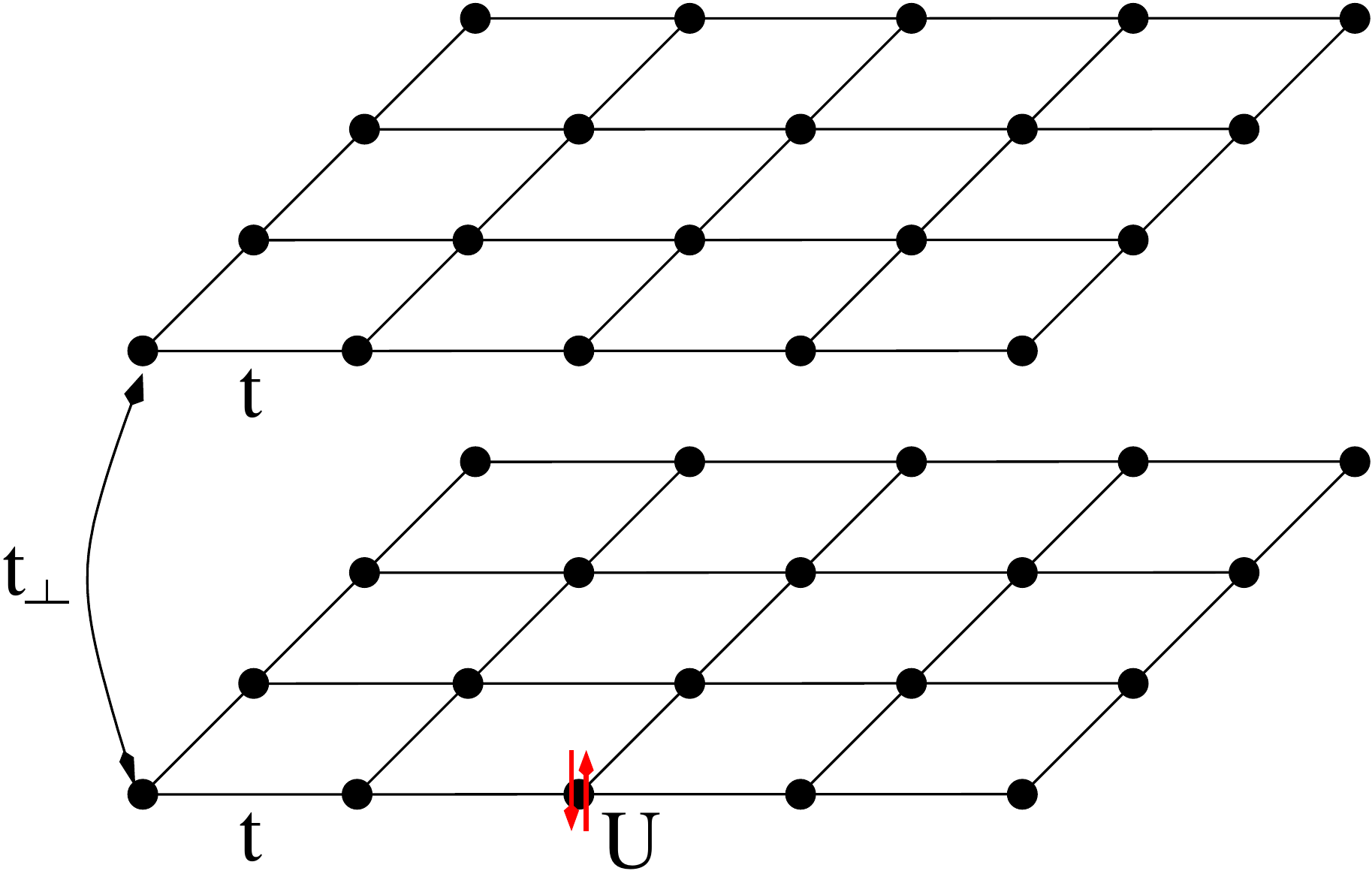}}
\hspace{3ex}
\includegraphics[height=0.35\columnwidth]{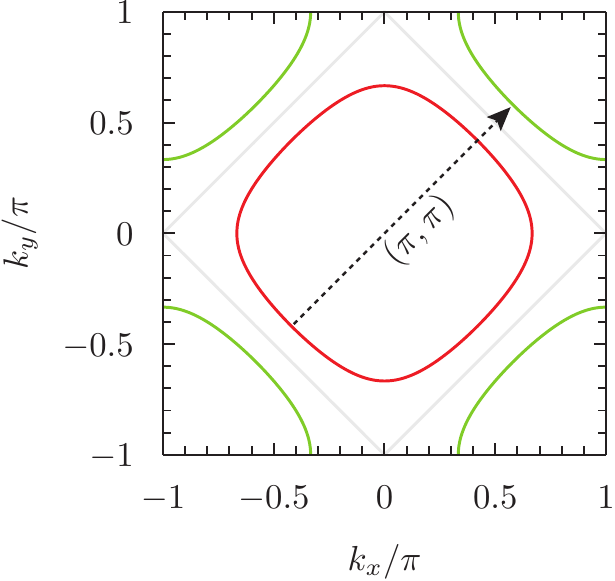}
\caption{Left: Illustration of the bilayer Hubbard model.
Right: Fermi surfaces of the bonding~(green) and
antibonding~(red) band for the square lattice bilayer with nearest neighbour
hopping for $t_\perp/t$=1. The symmetry of the dispersion relation ensures a perfect Fermi
surface nesting between the bands as long as~$t_\perp < 4$. The grey curve
is the Fermi surface for~$t_\perp=0$.}
\label{fig:2lfs}
\end{figure}

The bilayer Hubbard model on the square lattice  is given by the
following Hamiltonian:
\begin{eqnarray}\label{eq:bilayer_hamiltonian}
\hat H &= \hat H_t + \hat H_{t_\perp} + \hat H_U \\
\label{eq:H_t} \hat H_t &= - t \sum_{l, \sigma} \sum_{\langle i j \rangle}
\left( \hat c^\dagger_{j, l, \sigma} \hat c_{i, l, \sigma}^{\phantom\dagger} +
\hc \right) \\ \label{eq:H_t_perp} \hat H_{t_\perp} &= - t_\perp \sum_{i,
\sigma} \left( \hat c^\dagger_{i, 2, \sigma} \hat c_{i, 1,
\sigma}^{\phantom\dagger} + \hc \right) \\
\hat H_U &= U \sum_{i,l} \hat n_{i, l, \uparrow} \hat n_{i,l,\downarrow}\,.
\end{eqnarray}
Here $\hat c^{\dagger}_{i,l,\sigma}$ ($\hat c^{\phantom{\dagger}}_{i,l,\sigma}$) denotes
the creation (annihilation) operator of one electron on site $i$ and plane~$l
\in \{ 1,2 \}$ with spin $\sigma \in \{\uparrow,\downarrow\}$, while $\hat n_{i,l,\sigma}=
\hat c^{\dagger}_{i,l,\sigma} \hat c^{\phantom{\dagger}}_{i,l,\sigma}$ is the electron
density.  $t$ is the nearest neighbour hopping parameter in the plane, $t_\perp$
denotes the hopping between planes and $U$ is the on-site Coulomb repulsion, see \fref{fig:2lfs}.
Note that in the following we measure all quantities in units of~$t$.
The~$U=0$ dispersion,
\begin{equation}
\epsilon_{\vec k}^{\pm} = - 2 t \big( \cos(k_x) + \cos(k_y) \big) \pm t_\perp,
\end{equation}
for the bonding/antibonding ($\pm$) bands is at half filling perfectly nested, i.e.
$\epsilon_{\vec k+\vec Q}^{\pm} =\epsilon_{\vec k}^{\mp}$ with~$\vec Q = (\pi, \pi)$, as
illustrated in \fref{fig:2lfs}. As a consequence, the ground state~\cite{PhysRevB.31.4403} is an 
ordered antiferromagnet 
 for any $U>0$, as long as a Fermi surface 
exists, which is the case for $t_\perp/t<4$ in the 
limit $U\to0$. 

In this work we analyse the magnetic phase diagram at $T=0$ as well as the
paramagnetic case, which we  investigate by suppressing the long-range magnetic ordering.
This latter investigation, while it is done at $T=0$,  it is relevant for predictions  at finite and 
small temperatures
where long-range magnetic order is absent 
as a consequence of the Mermin-Wagner 
theorem~\cite{PhysRevLett.17.1133}.

In the limit of large interaction strength, 
$U \rightarrow \infty$, the Hubbard Hamiltonian in
equation~\eref{eq:bilayer_hamiltonian} reduces to the Heisenberg Hamiltonian,
\begin{equation}
\hat H = J \sum_l \sum_{\langle i j \rangle} \hat{ \vec S }_{i,l} \cdot \hat{
\vec S }_{j,l} + J_\perp \sum_i \hat{ \vec S }_{i,1} \cdot \hat{ \vec S
}_{i,2}\,, \\ 
\end{equation}
with $J = {4t^2}/{U}$ and $J_\perp = {4t_\perp^2}/{U}$.
The bilayer Heisenberg Hamiltonian has itself been subject of extensive
research, and it has been found to undergo an order-disorder transition
\cite{gros1995transition}
at~$J_\perp/J = 2.552$, which corresponds to~$t_\perp/t = 1.588$ in terms of
the Hubbard Hamiltonian's hopping
parameters~\cite{PhysRevLett.72.2777}\cite{PhysRevB.73.014431}.

Studies of the complete phase diagram of the bilayer Hubbard model have been
conducted by Fuhrmann~\etal using DMFT~\cite{PhysRevB.73.245118} and by
Kancharla and Okamoto with cluster DMFT~\cite{PhysRevB.75.193103}. 
The authors of reference~\cite{PhysRevB.73.245118} concentrated on the paramagnetic phase at finite temperature
and  found a metallic phase at small $U$ and small $t_\perp$ values as well as 
an insulating phase as $t_\perp$ increases (the band insulator) or
$U$ increases (the Mott insulator), but no clear 
separation between the Mott and the band insulating phase was found.

The authors of reference~\cite{PhysRevB.75.193103} considered clusters of sizes $2\times 2$
(i.e. two sites per plane)  within cluster DMFT and  
 performed  exact diagonalization calculations to solve the impurity problem, 
which allowed them to investigate $T=0$ and to obtain both a magnetic 
phase diagram as well as a~$T=0$ phase diagram where magnetism is
suppressed. These results show  several important differences with respect
to the earlier phase diagram by Fuhrmann~\etal~\cite{PhysRevB.73.245118}: 
First of all,  Kancharla \etal~\cite{PhysRevB.75.193103} distinguish between a Mott and 
a band insulator by looking at the behaviour of
the charge gap as a function of~$t_\perp$. Also, 
they find a Mott insulator at~$t_\perp = 0$ for any 
value of~$U > 0$, even if magnetism is suppressed. They mention 
the cluster DMFT's intralayer spatial correlation as the reason for this 
being found with cluster DMFT but not in the DMFT results by Fuhrmann~\etal~\cite{PhysRevB.73.245118}. 
However, at~$t_\perp = 0$ the planes are completely decoupled and 
should have the same properties as a single plane. For the single
plane, different methods predict that the system becomes a Mott insulator at some
finite~$U$, as shown for instance in reference~\cite{PhysRevLett.110.216405}, where the critical~$U$ 
was calculated using the dynamical cluster approximation with 
a variety of cluster sizes. 
Also, Tocchio~\etal~\cite{tocchio2012strong} and Capello~\etal~\cite{PhysRevB.73.245116} showed with the VMC method (which
in the first reference includes Fermi-surface renormalization effects) the appearance of a Mott insulator 
at a finite $U$.

A possible reason for the occurrence of a Mott insulator for any
finite~$U$  
in reference~\cite{PhysRevB.75.193103} might be the small cluster 
size of~$2 \times 2$  used in cluster DMFT. 
Having two sites in each plane breaks the fourfold 
rotational symmetry of the square lattice and results in an
artificially enhanced local pair within each plane for any~$U>0$. This can
noticeably affect the phase diagram of a system, as reported by Lee~\etal for
the two-orbital Hubbard model~\cite{PhysRevLett.104.026402}. A particularly
interesting feature of the phase diagram in reference~\cite{PhysRevB.75.193103} is
that for a certain range of~$U$ values the system goes through two phase transitions
as~$t_\perp$ is increased, first from a Mott insulator to a metal and then from
a metal to a band insulator, while at large~$U$ the system exhibits a direct
transition from a Mott to a band insulator. Analogous features have been also proposed 
in reference~\cite{PhysRevLett.97.046403} for the ionic Hubbard model. If magnetic order is allowed, it is
noteworthy that no magnetic ordering was found in the  cluster DMFT~\cite{PhysRevB.75.193103} for
small~$U$ and~$2 \lesssim t_\perp < 4$, even though there is a perfect nesting
between the Fermi surfaces of the bonding and antibonding bands 
with~$\vec Q = (\pi, \pi)$. A metallic phase at small $U$ in the magnetic phase diagram 
has been also obtained in a Determinant Quantum Monte Carlo (DQMC) study~\cite{PhysRevB.77.144527}. 
This result could be a consequence of the finite temperatures that 
have been used in DQMC, since they could 
be large enough to destroy the tiny magnetic order at small $U$, where the gap is exponentially small.

Overall, DMFT and previous DQMC results do not give a conclusive picture yet, making it
worthwhile to also employ other methods in order to 
gain a deeper understanding of the physics in the bilayer Hubbard
model.

\section{Methods}

The variational Monte Carlo method was introduced by McMillan in 1965 to
calculate the ground state of liquid $^4$He~\cite{PhysRev.138.A442}, and in
1977 applied to a fermionic system for the first time~\cite{PhysRevB.16.3081}.
Its basic idea is to use the Rayleigh-Ritz principle~\cite{Ritz09} to
approximate the ground state through a variational many-body wavefunction. It
is a Monte Carlo method because a stochastic sampling is used to evaluate the
sum over a high dimensional configuration space. A detailed description of how
the variational Monte Carlo method can be applied to the Hubbard model may be
found for instance in reference~\cite{Rue13}. The VMC approach has played
also a central role when examining the large-$U$ limit of the Hubbard
model~\cite{gros1989physics,valenti1992luttinger,edegger2007gutzwiller}, 
in the context of high-temperature superconductivity.

The choice of the variational many-body wavefunction is crucial in
order to obtain reliable results.
 Here, we define a variational
state~$\ket \Psi$ that consists of two parts: A Slater determinant~$\ket \Phi$
and  a Jastrow correlator~$\hat{\mathcal P}_J$ acting on $\ket \Phi$: \begin{equation}
\ket \Psi = \hat{ \mathcal P }_J \ket{\Phi}\,.
\end{equation}
Here the Slater determinant~$\ket \Phi$ ensures the antisymmetry of the
wavefunction while the Jastrow factor~$\hat{\mathcal P}_J$ modifies its
amplitude to take into account electronic correlations. 

The state~$\ket \Phi$ is the ground state of a variational mean-field
Hamiltonian~$\hat H_\mathrm{var}$ which may include up to five different terms:
Nearest neighbour hopping within the planes, hopping between the planes,
superconducting pairing in the planes with $d$-wave symmetry,  pairing between
the planes and an antiferromagnetic term, according to the following
equations~\cite{PhysRevB.38.931, 0953-2048-1-1-009, PhysRevB.45.5577}: 
\begin{eqnarray}
\hat H_\mathrm{var} &= \hat H_t + \hat H^\mathrm{(var)}_{t_\perp} + \hat
H_\Delta + \hat H_{\Delta_\perp} + \hat H_\mathrm{mag} \\ \
\label{2lay:eq:H_t} \hat H_t &= -t \sum_{\vec r, l} \left( \hat c^\dagger_{\vec
r + \vec e_x, l} + \hat c^\dagger_{\vec r - \vec e_x, l} + \hat c^\dagger_{\vec
r + \vec e_y, l} + \hat c^\dagger_{\vec r - \vec e_y, l} \right) \hat c_{\vec
r, l} \\ \
\label{2lay:eq:H_t_perp} \hat H_{t_\perp}^\mathrm{(var)} &= -
t_\perp^\mathrm{(var)} \sum_{\vec r} \left( \hat c^\dagger_{\vec r, 2} \hat
c_{\vec r, 1}^{\phantom\dagger} + \hc \right) \\ \
\hat H_\Delta &= \Delta \sum_{\vec r, l} \hat c^\dagger_{\vec r, l, \uparrow}
\Big( \hat c^\dagger_{\vec r + \vec e_x, l, \downarrow} + \hat c^\dagger_{\vec
r - \vec e_x, l, \downarrow} - \hat c^\dagger_{\vec r + \vec e_y, l,
\downarrow} - \hat c^\dagger_{\vec r - \vec e_y, l, \downarrow} \Big) + \hc \\
\
\hat H_{\Delta_\perp} &= \Delta_\perp \sum_{\vec r} \left( \hat c^\dagger_{\vec
r, 1, \uparrow} \hat c^\dagger_{\vec r, 2, \downarrow} + \hat c^\dagger_{\vec
r, 2, \uparrow} \hat c^\dagger_{\vec r, 1, \downarrow} + \hc \right) \\ \
\label{2lay:eq:H_mag} \hat H_\mathrm{mag} &= \mu_m \sum_{i} (-1)^{\tau(i)} \hat S^z_i\,.
\end{eqnarray}
Here $\vec r$ labels the sites within the planes, $l$ is the plane index, $\vec
e_{x (y)}$ is the unity vector along the $x (y)$ direction and $S^z_i$ indicates
the $z$ component of the spin operator on site $i$. Note that the square
lattice bilayer model is a bipartite lattice, with $\tau(i) \in \{ 1, 2 \}$ labelling
the sublattice of site~$i$, so that a different spin orientation is preferred
for each of the two sublattices when~$\mu_m > 0$. We would like to mention that
the following particle-hole transformation has been used in order to
diagonalize the variational Hamiltonian: \begin{eqnarray}
\label{eq:particle_hole_trans_up}\hat c_{i \uparrow} = \hat d_{i \uparrow}
\qquad &\mathrm{and} \qquad \hat c_{i \uparrow}^\dagger = \hat d_{i
\uparrow}^\dagger \\ \label{eq:particle_hole_trans_down}\hat c_{i \downarrow} =
\hat d_{i \downarrow}^\dagger \qquad &\mathrm{and} \qquad \hat c_{i
\downarrow}^\dagger = \hat d_{i \downarrow}\,.
\end{eqnarray}
This is possible because we chose the spins to align along the $z$ direction 
in the antiferromagnetic term of Eq.~(\ref{2lay:eq:H_mag}).

The alternative choice of aligning the spins along the $x$ direction, 
see for instance reference~\cite{PhysRevB.78.041101}, 
does not allow to study magnetism and pairing together as a single Slater determinant, but it is 
often preferred because the variational state is improved by the application of a spin-Jastrow factor 
$J_s=\exp\left( \frac{1}{2} \sum_{ij}u_{ij}S^z_iS^z_j\right)$ that 
couples spins in a direction orthogonal to the ordering one, see reference~\cite{PhysRevB.62.12700}. 

The Jastrow factor~$\hat{\mathcal P}_J$ implements a long-range density-density
correlation which has been shown to be essential in the variational description of
Mott insulators~\cite{PhysRevLett.94.026406}. In order to account for the
bilayer nature of the system we used a modified version of the Jastrow factor
with a different set of variational parameters for intraplane ($v$) and
interplane~($v_\perp$) correlations: \begin{equation}\label{eq:Jastrow}
\hat{\mathcal P}_J = \exp \left( \frac{1}{2} \sum_{ij} \sum_{l_1,l_2} \left[
v(ij) \delta_{l_1,l_2} + v_\perp(ij) (1 - \delta_{l_1,l_2} ) \right] \; \hat
n_{i,l_1} \hat n_{j, l_2} \right)\,, \end{equation}
with the $v(ij)$ and the $v_\perp(ij)$ being optimized independently for every
distance $|\vec r_i - \vec r_j|$.

The Metropolis algorithm~\cite{Met53} with single particle updates has been
used to generate the electronic configurations, while the optimization of the
variational wavefunction was done using the stochastic reconfiguration
method~\cite{PhysRevB.64.024512, casula:7110}, that allows us to independently
optimize every variational parameter in~$\hat{\mathcal P}_J$, as well as~$
t_\perp^{\mathrm{(var)}}$, $\Delta$, $\Delta_\perp$ and $\mu_m$ in the
mean-field state. The in-plane hopping parameter $t$ is kept fixed to set the
energy scale. A lattice size of $10\times10$ sites per plane was used, unless stated otherwise.

\section{Results}

\begin{figure}
\centering
\includegraphics[width=0.95\textwidth]{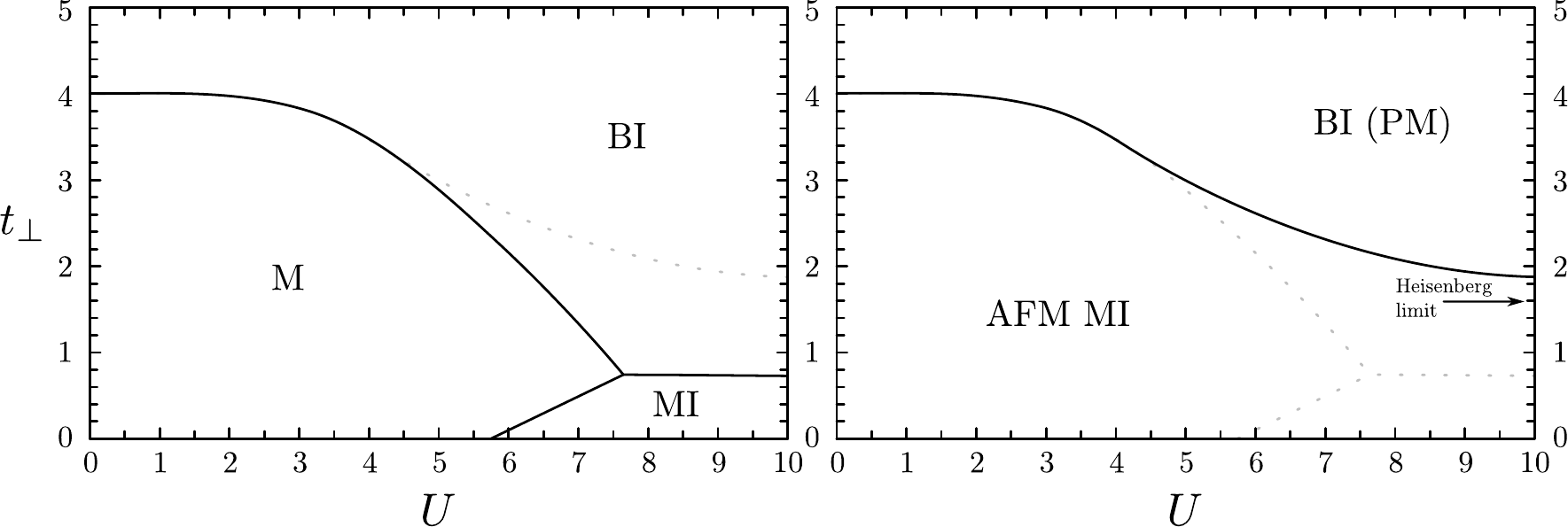}
\caption{Phase diagrams obtained with variational Monte Carlo for the square
lattice bilayer Hubbard model. The non-magnetic phase diagram (left) shows a
metallic phase~(M), a band insulating phase~(BI) and a Mott insulating
phase~(MI). The magnetic phase diagram (right) shows an antiferromagnetic Mott
insulator~(AFM MI) and a paramagnetic band insulator~(BI (PM)).} 
\label{fig:phase_diagram}
\end{figure}

The non-magnetic and the magnetic phase diagrams obtained using
VMC simulations are presented in \fref{fig:phase_diagram}.

The non-magnetic phase diagram (\fref{fig:phase_diagram}, left panel)
shows a metallic phase at small~$U$
and~$t_\perp$, which goes, as expected, into a band insulating 
phase at~$t_\perp = 4$. The 
critical value of~$t_\perp$ that is needed to make the system band insulating 
decreases as~$U$ is increased. At small~$t_\perp$ the system undergoes a metal
to Mott insulator transition as~$U$ is increased, with the critical~$U$ ranging
from about~$5.5$ at~$t_\perp = 0$ to~$7.5$ at~$t_\perp = 0.7$. For large~$U$
the system undergoes a Mott to band insulator transition at~$t_\perp \approx
0.7$, where the critical value of~$t_\perp$ is independent of~$U$ and only
weakly dependent on the system's size. Hysteresis in the variational parameters
when going through the Mott to band insulator transition suggests the
transition to be of first order. An interesting feature is that for~$5.5
\lesssim U \lesssim 7.5$ the system first undergoes a Mott insulator to metal
transition and then a metal to band-insulator transition as $t_\perp$ is
increased.

In contrast, only two phases are found in the magnetic phase diagram
(\fref{fig:phase_diagram}, right panel). The system is a
N\'{e}el ordered antiferromagnetic Mott insulator as long as~$t_\perp$ is
smaller than some critical value, and a paramagnetic band insulator for
larger~$t_\perp$. The critical interlayer hopping is~$t_\perp = 4$ at~$U=0$ and
decreases as~$U$ is increased. At~$U=10$ it reaches a value of about~$t_\perp=1.9$,
which suggests that the critical interlayer hopping approaches the Heisenberg limit of~$t_\perp=1.588$ for~$U
\rightarrow \infty$. We now discuss the  details of the calculations
undertaken in order to obtain the above presented phase diagrams.

\subsection{The non-magnetic phase diagram}

In order to obtain the non-magnetic phase diagram it is useful to first
analytically diagonalize the variational single-particle Hamiltonian whose
eigenstates are used to build the determinantal part of the wavefunction. The
argument here is that if the Slater determinant is already gapped, no
correlator can make the wavefunction conducting again. Therefore, one can identify a
band insulator purely by looking for a gap in the band structure of the
determinantal part. Note that also a superconductor has a gap in the mean field
state, without being an insulator. However, the only parameter in our
variational wave function that could induce superconductivity is the in-plane
pairing~$\Delta$ that is always gapless in the nodal direction, since it has
$d$-wave symmetry. Therefore, we do not risk to accidentally classify a
superconductor as an insulator, and the existence of a gap in the mean field
state is indeed equivalent to the system being insulating.

\begin{figure}
\centering
\includegraphics[width=0.75\textwidth]{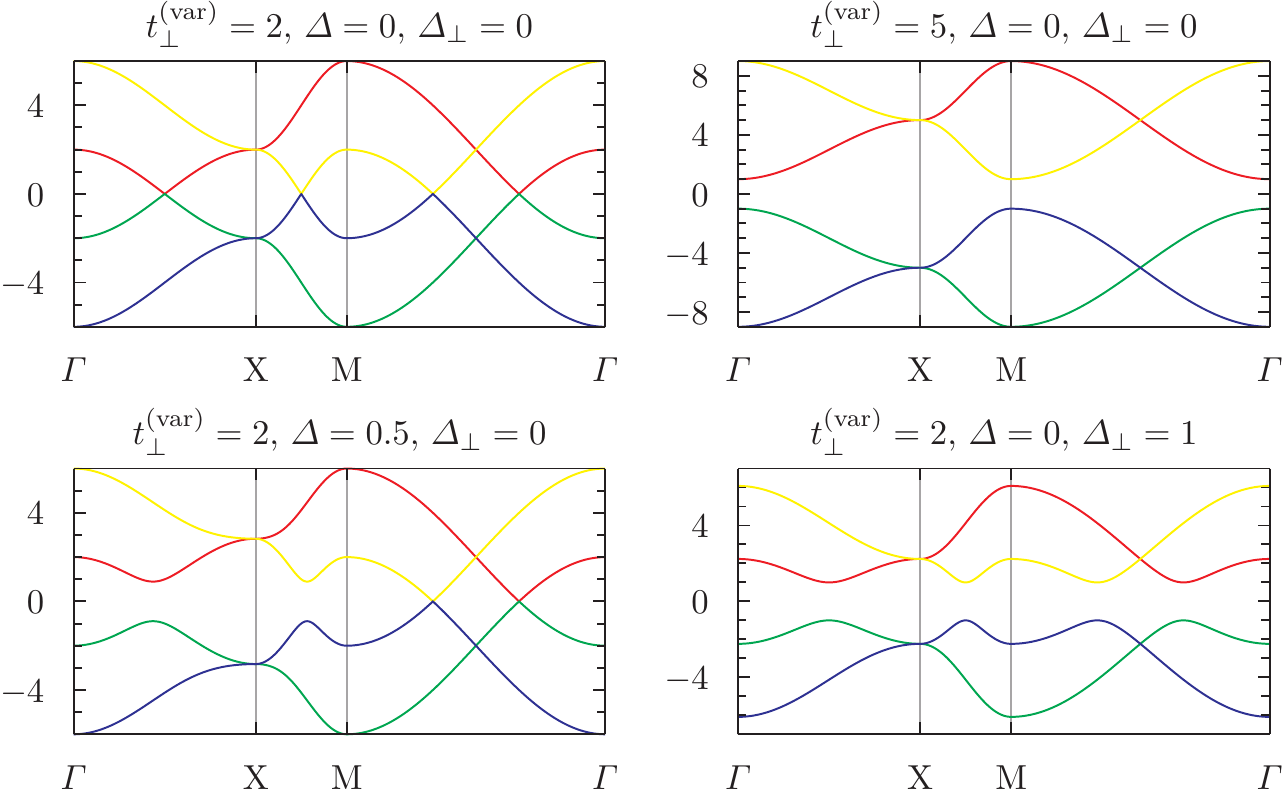}
\caption{Band structure of the Slater determinant,
compare equation~\eref{eq_13} and \eref{eq_24}, for
different values of the variational parameters. The symmetry points are~$\Gamma
\equiv (0,0)$, $X \equiv (\pi,0)$ and $M \equiv (\pi, \pi)$. The blue and green
band are occupied, and a gap opens for~$t^\mathrm{(var)}_\perp > 4$
or~$\Delta_\perp \neq 0$. As expected, the in-plane pairing~$\Delta$ does not
open a gap in the $d$-wave's nodal direction~$\Gamma \rightarrow M$.}
\label{fig:nonmag_bands}
\end{figure}

In the non-magnetic case the variational Hamiltonian consists of four terms for
the intraplane/interplane hopping and pairing: \begin{equation}
\hat H_\mathrm{var} = \hat H_t + \hat H^\mathrm{(var)}_{t_\perp} + \hat
H_\Delta + \hat H_{\Delta_\perp}\,.  \end{equation}
Diagonalizing it analytically using the particle-hole transformation gives the
following bandstructure: 
\begin{eqnarray}
\label{eq_13}
\fl \epsilon_{\vec k}^{(1,3)} &= \pm \sqrt{ \big( -2 t ( \cos k_x + \cos k_y )
+ t^\mathrm{(var)}_\perp \big)^2 + \big( 2 \Delta ( \cos k_x - \cos k_y ) +
\Delta_\perp \big)^2} \\
\fl \epsilon_{\vec k}^{(2,4)} &= \pm \sqrt{ \big( -2 t ( \cos k_x + \cos k_y )
- t^\mathrm{(var)}_\perp \big)^2 + \big( 2 \Delta ( \cos k_x - \cos k_y ) -
\Delta_\perp \big)^2}\,. 
\label{eq_24}
\end{eqnarray}
There are two bands with only positive/negative energies. At half
filling there are enough electrons to populate exactly two bands, so that the
two negative bands are always completely filled and the two positive bands
entirely empty. Consequently, the only way not to have a bandgap is by having
the bands touch each other at zero energy. The easiest way to understand the
influence of the different parameters is to look at \fref{fig:nonmag_bands}, 
where the bands are plotted for different values of the variational parameters.

It can easily be seen that there are two ways of opening a gap in the mean
field part of our variational wavefunction, namely having
a~$t^\mathrm{(var)}_\perp > 4$, or a non-zero~$\Delta_\perp$ which corresponds
to the formation of singlets between the planes. We now discuss
the variational Monte Carlo simulation results used to draw
the non-magnetic phase diagram in \fref{fig:phase_diagram}.

\begin{figure}
\centering
\includegraphics[width=0.75\columnwidth]{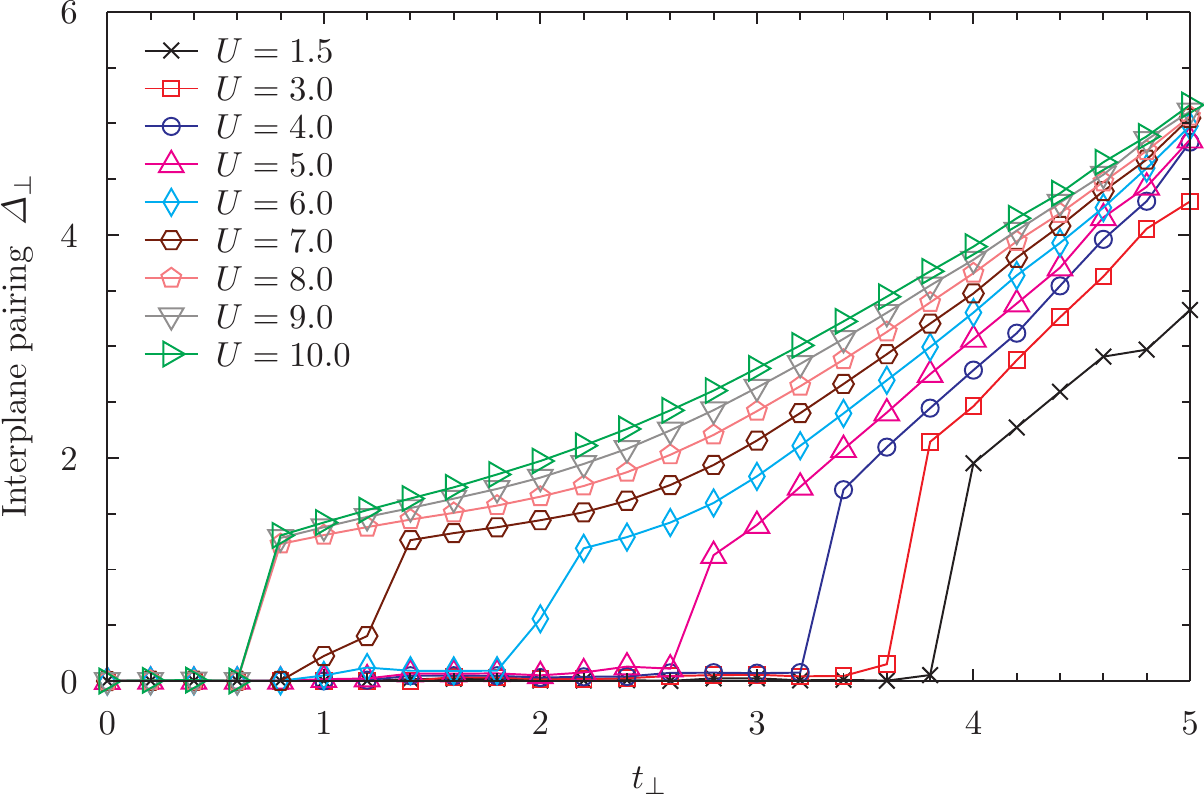}
\caption{Interplane pairing~$\Delta_\perp$ as a function of the
interplane hopping~$t_\perp$   for the non-magnetic phase diagram of \fref{fig:phase_diagram}.
The system becomes a band insulator
for $\Delta_\perp\ne0$ (compare with \fref{fig:nonmag_bands}).
} 
\label{fig:nonmag_Deltaperp}
\end{figure}

\Fref{fig:nonmag_Deltaperp} shows the optimized interplane
pairing~$\Delta_\perp$ as a function of the interplane hopping~$t_\perp$.

Starting with~$t_\perp = 4$ at~$U = 0$ the region with a non-zero~$\Delta_\perp$
extends to lower~$t_\perp$ as~$U$ is increased. For~$U \geq 8$ the jump is at a
constant~$t_\perp \approx 0.7$. As any non-zero~$\Delta_\perp$ opens a gap in
the mean field part of the wavefunction, the region of a non-zero~$\Delta_\perp$
maps out the band insulating part of the phase diagram in
\fref{fig:phase_diagram}.

\begin{figure} 
\centering
\includegraphics[width=0.75\columnwidth]{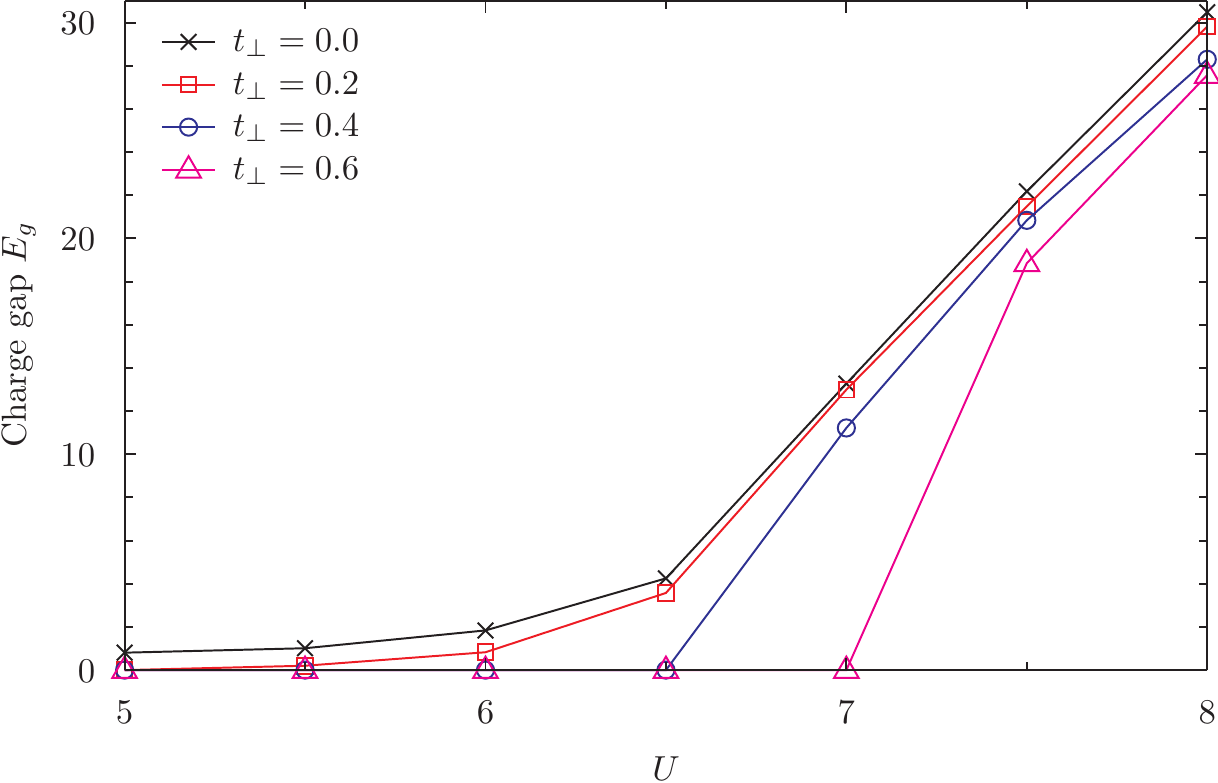}
\caption{ Charge gap~$E_g$ for the non-magnetic phase diagram 
(in arbitrary units, compare equation~\ref{eq_E_g}) as a function of 
the Coulomb repulsion~$U$.}
\label{fig:nonmag_mottgap}
\end{figure}

At variance with the band insulator, a Mott insulating region is
characterized by a gapless mean-field state, while the insulating nature is
driven by the electronic correlations that are included in the Jastrow
factor~$\hat{\mathcal P}_J$. In order to discriminate between a Mott insulator
and a metal, we use the following single mode ansatz for the wavefunction of
the excited state, which goes back to  Richard Feynman's work on excitations
in liquid Helium~\cite{PhysRev.94.262}, and was later successfully applied to
fermionic systems:~\cite{PhysRevB.3.1888}\cite{PhysRevB.33.2481}
\begin{equation}
\ket{\Psi_{\vec q}} = \hat n_{\vec q} \ket{\Psi_0}\,, 
\end{equation}
where $\hat n_{\vec q}$ is the Fourier transform of the particle density with $\vec q=(q_x,q_y,0)$. 
By calculating the energy of the excited state, one can derive a formula for an
upper bound of the charge gap~$E_g$, that relates it to the static structure
factor~$N(\vec q) = \langle \hat n_{-\vec q} \; \hat n_{\vec q}
\rangle$:~\cite{PhysRevB.83.195138} 
\begin{equation}
E_g \propto \lim_{\vec q \rightarrow 0} \frac{|\vec q|^2}{N(\vec q)}\,.
\label{eq_E_g}
\end{equation}
\Fref{fig:nonmag_mottgap} shows the charge gap~$E_g$ as a function of
the Coulomb repulsion~$U$ for different interplane hoppings~$t_\perp$.

As expected from the known monolayer results, we find the system to be a Mott
insulator for large enough values of~$U$. Note that the critical~$U$ needed to
make the system Mott insulating, i.e. when the charge gap $E_g$ starts to grow as a function of $U$, 
increases from about~$U=5.5$ at~$t_\perp = 0$
to~$U=7.5$ at~$t_\perp = 0.6$. Note that a finite charge gap for $U\lesssim 5.5$ and $t_{\perp}=0$ 
is just an artefact of the limited number of $\vec q$ points that are available for the extrapolation to $\vec q=0$ 
and indeed decreases as the system size increases.  
We point out that a sizeable in-plane~$\Delta$ can
be found only in the Mott insulating region, indicating that the pairing within
the planes is to be understood in terms of the resonating valence bond
theory~\cite{ANDERSON06031987,gros1988superconductivity}, in which $d$-wave pairs 
are formed, but not phase coherent. Indeed, it is the presence of the Jastrow factor of equation~\eref{eq:Jastrow}, 
that allows the $\hat{{\mathcal P}}_J\ket{\Phi}$ wave function to describe a Mott insulator 
through the opening of a charge gap without any symmetry breaking. 

\begin{figure}
\centering
\includegraphics[width=0.75\columnwidth]{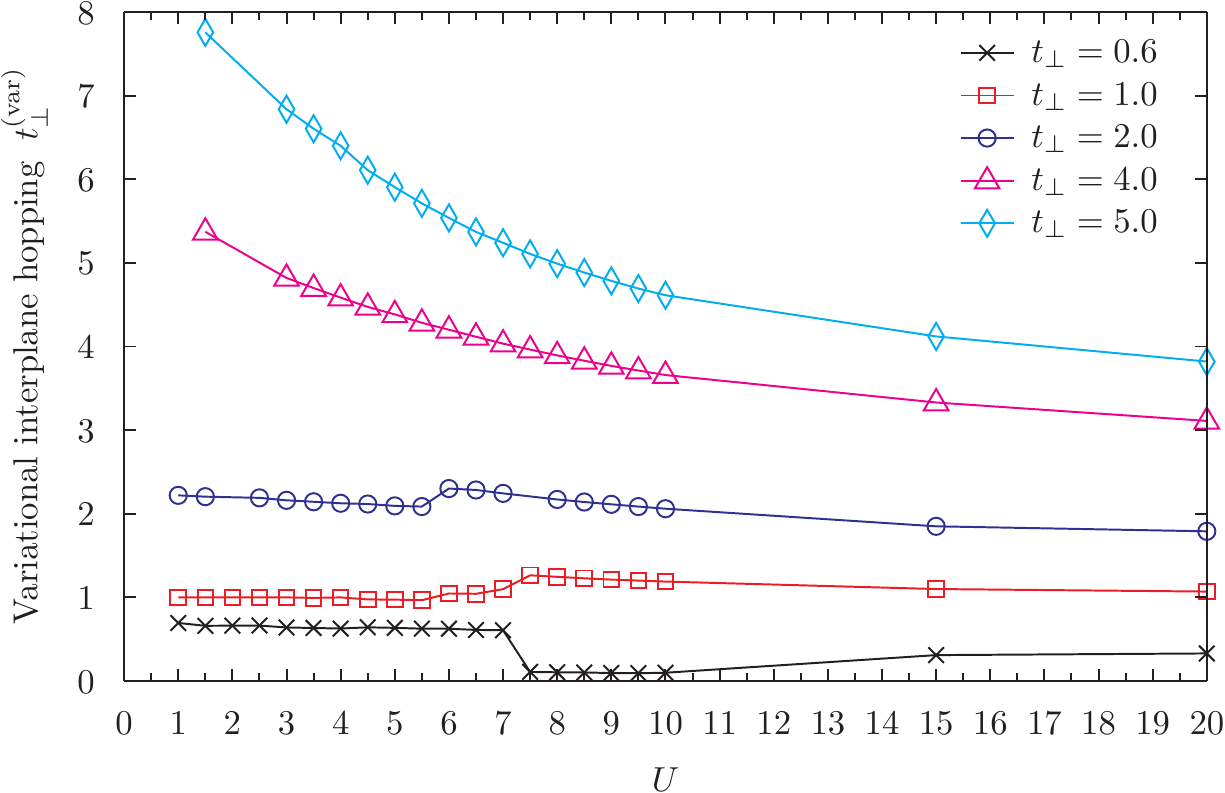}
\caption{The variational interplane hopping~$t^\mathrm{(var)}_\perp$ 
as a function of the Coulomb repulsion~$U$ for different values 
of the interplane hopping~$t_\perp$ in the original Hubbard 
Hamiltonian. The results are for the non-magnetic phase diagram.}
\label{fig:nonmag_t_perp_renorm}
\end{figure}

\Fref{fig:nonmag_t_perp_renorm} shows the renormalization of the variational
interplane hopping~$t^\mathrm{(var)}_\perp$. 

In general the variational~$t^\mathrm{(var)}_\perp$ is not too different from
the~$t_\perp$ in the original Hubbard Hamiltonian, but there are two exceptions
to this rule. The first one is that in the Mott insulating phase (at large~$U$
and small~$t_\perp$) the variational~$t^\mathrm{(var)}_\perp$ is renormalized
to quite small values. Together with the lack of a~$\Delta_\perp$ in this
region, this indicates that the mean field part of the wavefunction is almost
that of two decoupled Hubbard planes. The other exception is that for
small~$U$ and large~$t_\perp$ the variational~$t^\mathrm{(var)}_\perp$ is
renormalized to values larger than the original~$t_\perp$. This is due to the fact
that in the limit $U\to0$ the 
  ground-state wavefunction is independent of the value 
of~$t_\perp$ with the bonding band filled as long as~$t_\perp>4$.

\begin{figure}
\centering
\includegraphics[width=0.75\columnwidth]{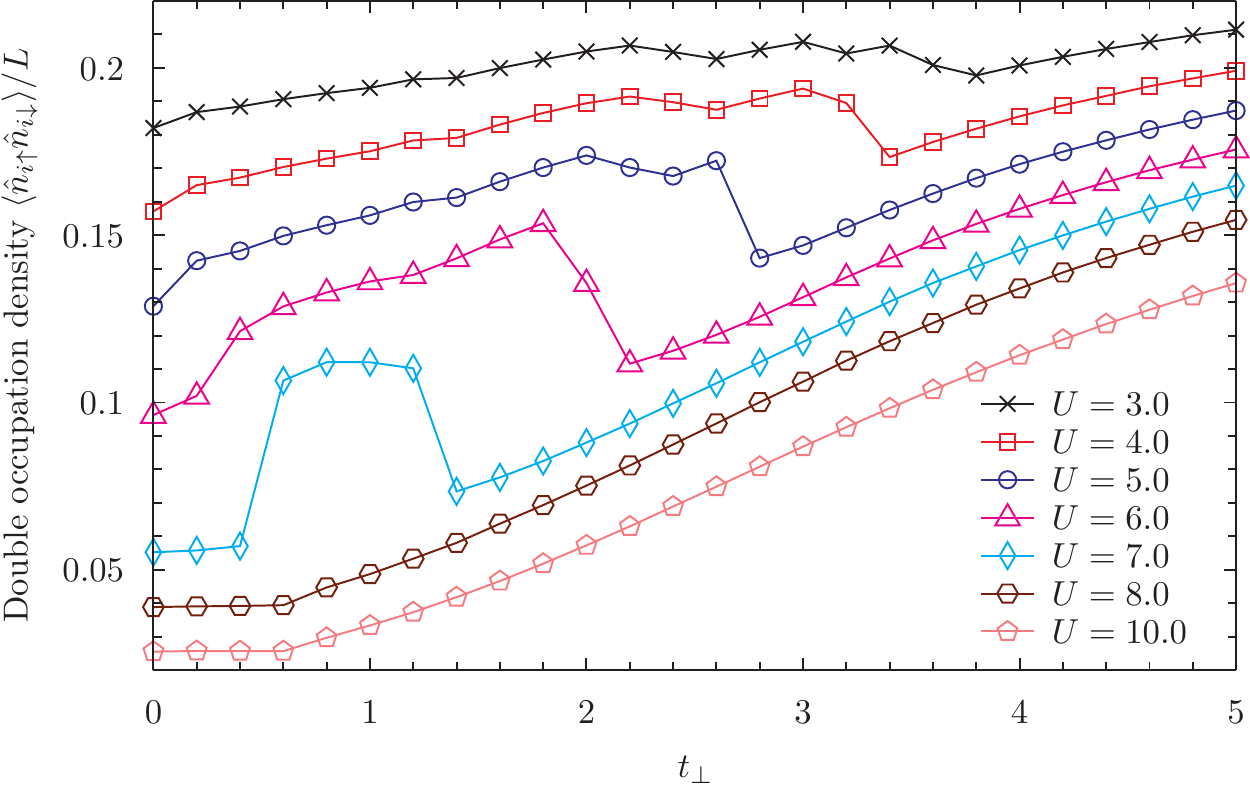}
\caption{For the non-magnetic phase diagram, \fref{fig:phase_diagram},
the density of double occupancies as a function of the 
interplane hopping~$t_\perp$.} 
\label{fig:nonmag_dblocc}
\end{figure}

The band gap and the Mott gap opening via~$\Delta_\perp$, 
presented in \fref{fig:nonmag_Deltaperp} and 
\fref{fig:nonmag_mottgap} respectively, 
suffice for mapping out the regions of band and 
Mott insulators in the phase diagram. The physics
of the respective transitions can be confirmed by 
looking at the density of double occupancies 
presented in \fref{fig:nonmag_dblocc},
as a function of the interplane hopping~$t_\perp$.

For intermediate values of~$U$, the density of double occupancies first rises
sharply and then drops off abruptly again as the~$t_\perp$ is increased,
signalling that there is a metallic phase in between the Mott and band
insulating phases for~$5.5 \lesssim U \lesssim 7.5$. The double occupation density 
 was also calculated within $2\times2$  cluster DMFT~\cite{PhysRevB.75.193103}, but,
contrary to our results of \fref{fig:nonmag_dblocc}, it was
found that it decreases in the metallic phase. We believe that
our results correspond to the physics of the Mott insulating phase,
which suppresses double occupancies.

\begin{figure}
\centering
\includegraphics[width=0.75\columnwidth]{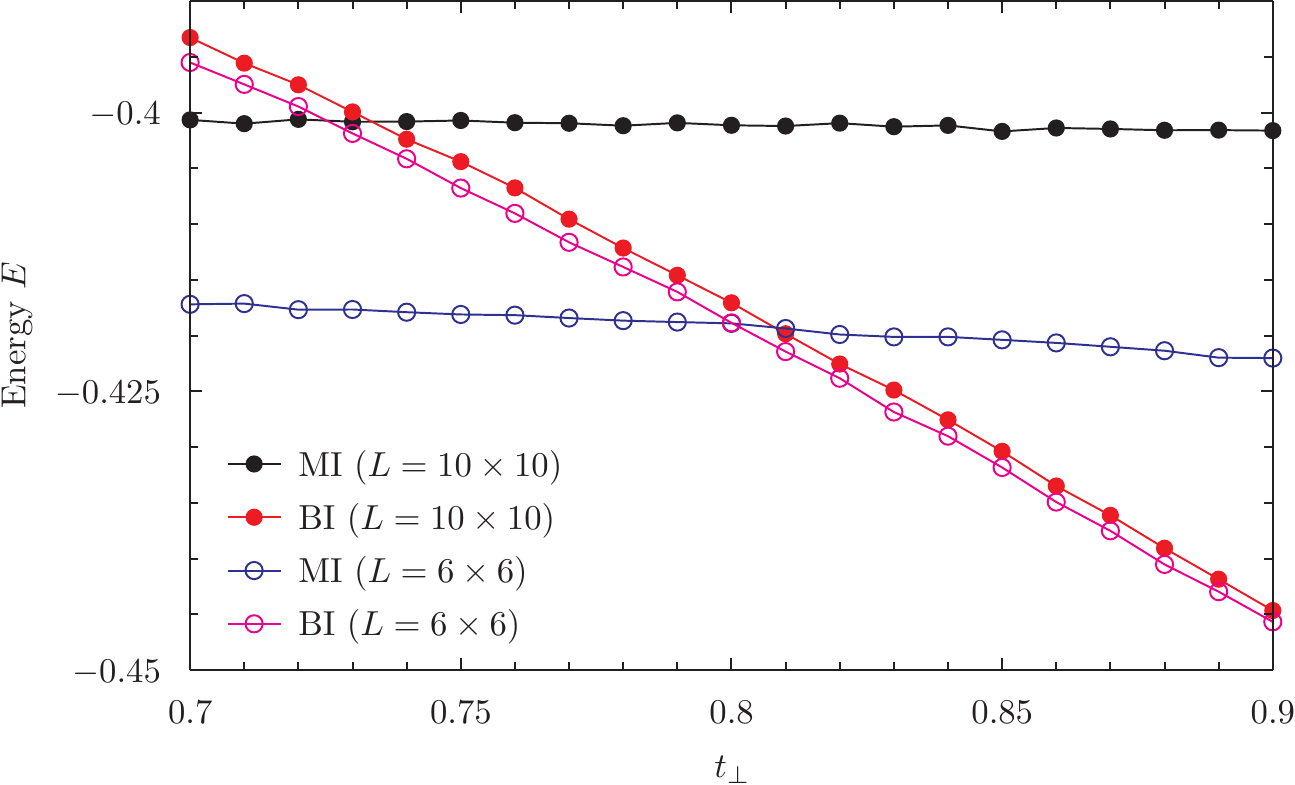}
\caption{The energy of both Mott and band insulating 
wavefunctions for different values of~$t_\perp$ close to the Mott to
band insulator transition. Note that increasing the system size from~$36$
to~$100$ sites per plane slightly decreases the critical~$t_\perp$ from~$0.8$
to~$0.73$. The data is for the non-magnetic phase diagram.} 
\label{fig:nonmag_MIBItrans}
\end{figure}

At large~$U \gtrsim 8$ the Mott insulator goes directly into the band insulator
at~$t_\perp \approx 0.7$. The Mott insulating wavefunction is characterized by
an in-plane~$\Delta > 0$ and a~$\Delta_\perp = 0$, while the band insulating
wavefunction has a sizeable~$\Delta_\perp$ but no in-plane~$\Delta$. A strong
hysteresis in these variational parameters was observed when going through the
transition, so that both a Mott and a band insulator could be obtained
for~$t_\perp$ close to the transition. The optimal wavefunction is the one with
the lowest energy, as plotted in \fref{fig:nonmag_MIBItrans}.
Finding hysteresis means that both the Mott and the band insulating
wavefunctions are local energy minima in our variational parameter space, thus
suggesting that the Mott to band insulator transition is of first
order. 

\subsection{The magnetic phase diagram}

The simulations through which the magnetic phase diagram was obtained were
performed in the same way as those for the non-magnetic case, except that the
site and spin dependent chemical potential~$\mu_m$ of equation~\eref{2lay:eq:H_mag} was no longer fixed 
to zero during the optimization.

\begin{figure}
\centering
\includegraphics[width=0.75\columnwidth]{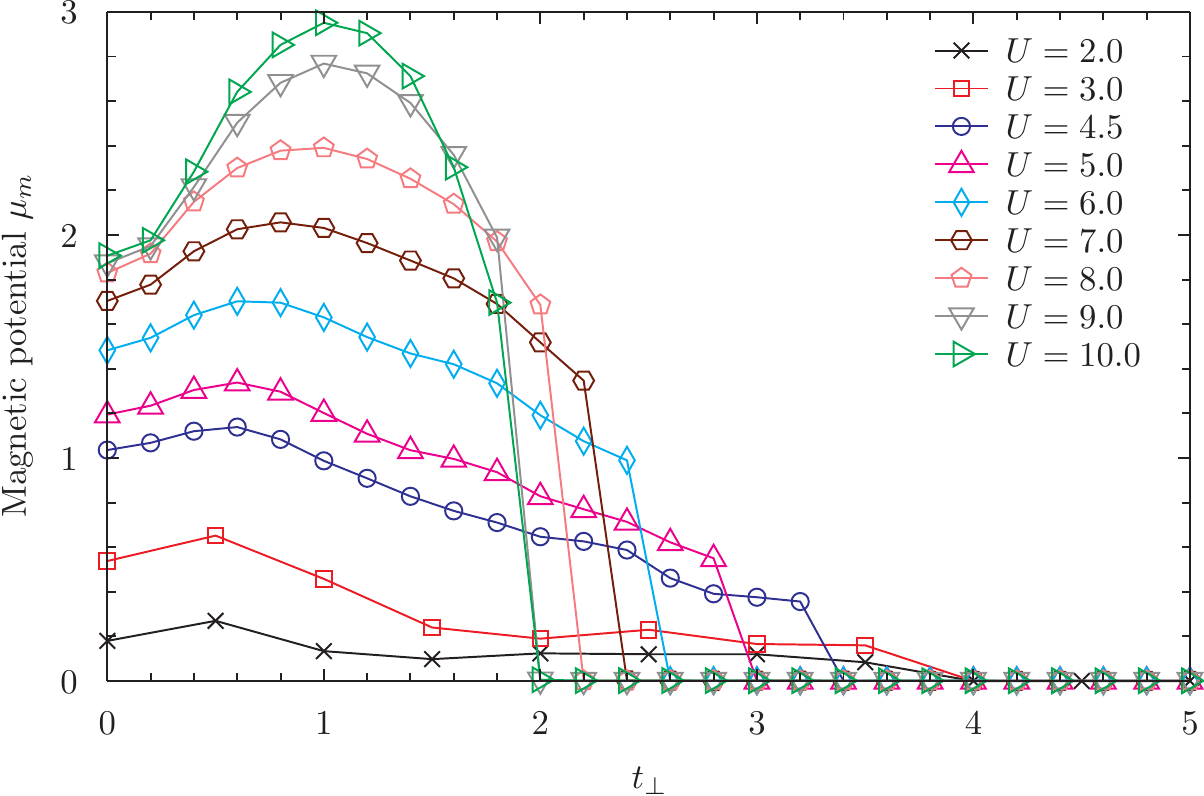}
\caption{The magnetic potential~$\mu_m$ as a function of
the interplane hopping~$t_\perp$.} 
\label{fig:mag_mu_m}
\end{figure}

\Fref{fig:mag_mu_m} shows the magnetic potential~$\mu_m$, 
as a function of the interplane hopping~$t_\perp$.

For small~$U$ an antiferromagnetic order is found for~$t_\perp < 4$, since
the ordering is driven by the perfect nesting of the Fermi surfaces corresponding
to partially filled bonding and antibonding
bands.  The Fermi surface  no longer exists if the bonding band is completely
filled and the antibonding band completely empty at~$t_\perp > 4$. Increasing
$U$ pushes the antiferromagnetic region to smaller~$t_\perp$, indicating
that the critical interplane hopping goes to the Heisenberg limit
of~$t_\perp = 1.588$ for~$U \rightarrow \infty$.

One can show, by analogy to the non-magnetic case in the previous section, that any
non-zero~$\mu_m$ makes the system insulating by analytically diagonalizing the
variational single particle Hamiltonian: \begin{equation}
\hat H_\mathrm{var} = \hat H_t + \hat H^\mathrm{(var)}_{t_\perp} + \hat
H_\mathrm{mag}\,, \end{equation}
which leads to the results
\begin{eqnarray}
\epsilon^{(1,3)}_{\vec k} &= \pm \sqrt{ \big( - 2 t ( \cos k_x + \cos k_y ) +
t_\perp \big)^2 + \mu_m^2 } \\ \epsilon^{(2,4)}_{\vec k} &= \pm \sqrt{ \big( -
2 t ( \cos k_x + \cos k_y ) - t_\perp \big)^2 + \mu_m^2 }\,.
\end{eqnarray}
The two negative bands are filled and the system can only be conducting if
these bands touch the empty bands at zero energy. Looking at the equations for the
energy bands, one can see that this can not happen for~$\mu_m \neq 0$, and hence
one can use a non-zero~$\mu_m$ as a criterion for an insulating state. Note that
we classify the ordered state in the phase diagram (\fref{fig:phase_diagram}, right panel) as a Mott insulator, 
even though we have a gap in the mean field state. This is due to the fact that the
antiferromagnetic ordering is correlation induced.

\begin{figure}
\centering
\includegraphics[width=0.75\columnwidth]{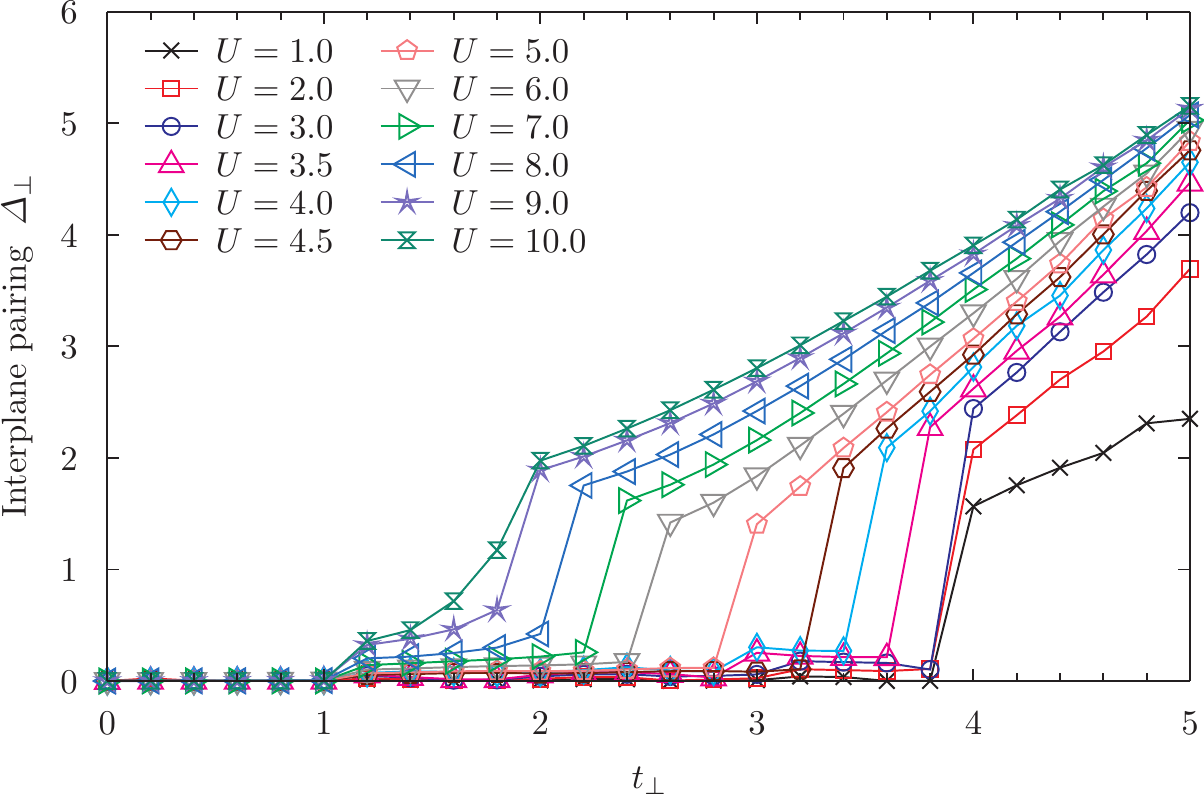}
\caption{The interplane pairing~$\Delta_\perp$ as a
function of the interplane hopping~$t_\perp$ for the magnetic 
phase diagram of \fref{fig:phase_diagram}.}
\label{fig:mag_Deltaperp}
\end{figure}

Comparing the interplane~$\Delta_\perp$ in \fref{fig:mag_Deltaperp}
 with the plot of the magnetic potential~$\mu_m$ in
\fref{fig:mag_mu_m}, we see that a non-zero~$\Delta_\perp$ is found in the
entire paramagnetic region. Both the~$\mu_m$ and the~$\Delta_\perp$ open a gap
in the mean field part of the wavefunction, so that the square lattice bilayer
Hubbard model is \emph{always} an insulator at~$U>0$ if magnetic order is
allowed.

It is interesting to note that at large~$U \gtrsim 8$ and~$1 < t_\perp \lesssim
2$ there is a small region with both a non-zero magnetic potential~$\mu_m$ and
an interplane pairing~$\Delta_\perp$. This, together with the fact that no
hysteresis in the variational parameters was observed at the order-disorder
transition, suggests that the transition is indeed  continuous.

\section{Conclusion}

In summary we have calculated the magnetic and non-magnetic phase diagram of
the square lattice bilayer Hubbard model using the variational Monte Carlo
method, as summarized in \fref{fig:phase_diagram}. Moreover, our results suggest that the Mott insulator 
to band-insulator transition is of first order in the non-magnetic phase diagram, while it becomes continuous 
when magnetic order is allowed.

Comparison of our results to the ones obtained with DMFT~\cite{PhysRevB.73.245118}
and cluster DMFT~\cite{PhysRevB.75.193103} reveals that our non-magnetic phase
diagram includes some features observed in DMFT and $2\times2$ cluster DMFT but also new distinct properties:
 While in agreement with $2\times2$ cluster DMFT there is a region in which the system first goes
from a Mott insulator to a metal and then to a band insulator as $t_\perp$
is increased, we do not find that this region extends down to~$U=0$. Instead
for the decoupled planes we find a metal to Mott insulator transition at a
critical~$U \approx 5.5$, which agrees with the DMFT results by 
Fuhrmann~\etal{}, and the dynamical cluster approximation results for the single layer by Gull
\etal~\cite{PhysRevLett.110.216405}. For large~$U$ our results agree with those
of $2\times2$ cluster DMFT in that there is a direct transition from a Mott to a
band insulator, but our critical~$t_\perp$ is smaller by about a factor of
3. The reason for this might be the cluster with two sites in each plane
used by the authors of reference~\cite{PhysRevB.75.193103} that breaks the fourfold rotational symmetry of
the square lattice and creates an artificially enhanced local pair within each
plane, ultimately stabilizing the in-plane Mott phase against the interplane
dimers of the band insulating phase.

The magnetic phase diagram, instead, shows clearly a different behaviour
from the one predicted by $2\times2$ cluster DMFT calculations.
The most obvious difference is that we no
longer find a metallic phase if magnetic ordering is allowed. Instead we find a
N\'{e}el ordered Mott insulator, which we attribute to the perfect nesting
between the bonding and antibonding band's Fermi surfaces. The reason for the
variational Monte Carlo approach stabilizing magnetic ordering compared to cluster DMFT
with two sites per plane might be the much more explicit treatment of long
range correlations.

The variational Monte Carlo results 
improve our understanding of the bilayer Hubbard model, but further
investigation may be necessary to clarify the origin of the sizeable
differences between VMC and DMFT results.

\ack

We would like to thank Federico Becca for useful discussions and the Deutsche Forschungsgemeinschaft for financial
support through grant SFB/TR~49.

\section*{References}
\bibliography{mypaper}{}

\end{document}